\documentclass[10pt,final,journal]{IEEEtran}
\IEEEoverridecommandlockouts

\usepackage{amsmath}
\usepackage{graphicx}
\usepackage{amssymb}
\usepackage{cases}
\usepackage{cite}
\usepackage[ruled,vlined,lined,ruled,linesnumbered]{algorithm2e}
\usepackage{relsize}
\usepackage[nodisplayskipstretch]{setspace}
\usepackage[usenames, dvipsnames]{color}
\usepackage{epstopdf}
\usepackage[font=small,skip=-5pt]{caption}
\usepackage{multirow}
\usepackage[utf8]{inputenc}
\usepackage{amsfonts}

\usepackage{booktabs}
\usepackage{bbm}
\usepackage{mathrsfs}

\newcommand{\Rmnum}[1]{\expandafter\@slowromancap\romannumeral #1@}
\usepackage{subfigure}
\usepackage{epsfig}
\usepackage[numbers,sort&compress]{natbib}
\usepackage{xcolor}
\usepackage{color}
\usepackage{enumerate}
\usepackage{extarrows}
\usepackage[ruled,linesnumbered,lined]{algorithm2e}  
\usepackage{algorithmic}
\usepackage{tensor}
\usepackage[colorlinks,linkcolor=blue,anchorcolor=blue,citecolor=blue,urlcolor=black]{hyperref}
\usepackage{url}

\begin{document}
\title{\LARGE{Neuromorphic Integrated Sensing and Communications}}
\author{Jiechen Chen, Nicolas Skatchkovsky, and Osvaldo Simeone \\
        \thanks{The authors are with the King’s Communications, Learning and Information Processing (KCLIP) lab, King’s College London, London, WC2R 2LS, UK. (email:\{jiechen.chen, nicolas.skatchkovsky, osvaldo.simeone\}@kcl.ac.uk).}
        \thanks{This work of Osvaldo Simeone and Nicolas Skatchkovsky was supported by the European Research Council (ERC) under the European Union’s Horizon 2020 Research and Innovation Programme (grant agreement No. 725731), and the work by Jiechen Chen was funded by the China Scholarship Council and King’s College London for their Joint Full-Scholarship (K-CSC) under Grant CSC202108440223.}\vspace*{-0.9cm}}

\maketitle

\begin{abstract}
Neuromorphic computing is an emerging technology that support event-driven data processing for applications requiring efficient online inference and/or control. Recent work has introduced the concept of neuromorphic communications, whereby neuromorphic computing is integrated with impulse radio (IR) transmission to implement low-energy and low-latency remote inference in wireless Internet-of-Things (IoT) networks. In this paper, we introduce neuromorphic integrated sensing and communications (N-ISAC), a novel solution that enables efficient online data decoding and radar sensing. N-ISAC leverages a common IR waveform for the dual purpose of conveying digital information and of detecting the presence or absence of a radar target. A spiking neural network (SNN) is deployed at the receiver to decode digital data and to detect the radar target using directly the received signal. The SNN operation is optimized by balancing performance metrics for data communications and radar sensing, highlighting synergies and trade-offs between the two applications.
\end{abstract}
\newtheorem{definition}{\underline{Definition}}[section]
\newtheorem{fact}{Fact}
\newtheorem{assumption}{Assumption}
\newtheorem{theorem}{\underline{Theorem}}[section]
\newtheorem{lemma}{\underline{Lemma}}[section]
\newtheorem{proposition}{\underline{Proposition}}[section]
\newtheorem{corollary}[proposition]{\underline{Corollary}}
\newtheorem{example}{\underline{Example}}[section]
\newtheorem{remark}{\underline{Remark}}[section]
\newcommand{\mv}[1]{\mbox{\boldmath{$ #1 $}}}
\newcommand{\mb}[1]{\mathbb{#1}}
\newcommand{\Myfrac}[2]{\ensuremath{#1\mathord{\left/\right.\kern-\nulldelimiterspace}#2}}
\newcommand\Perms[2]{\tensor[^{#2}]P{_{#1}}}
\newcommand{\note}[1]{[\textcolor{red}{\textit{#1}}]}

\vspace{-10pt}

\section{Introduction}
\vspace{-3pt}
Integrated sensing and communications (ISAC), a key enabling technology for 6G systems, leverages shared radio resources and hardware to realize the functions of sensing and communication. ISAC can enhance energy and spectral efficiencies by supporting context-aware decision making, whereby wireless devices, such as mobile phones and vehicles, act based on information about their surrounding environment \cite{liu2020joint}.

\begin{figure}[t!]
	\centering
	\includegraphics[width=3.5in]{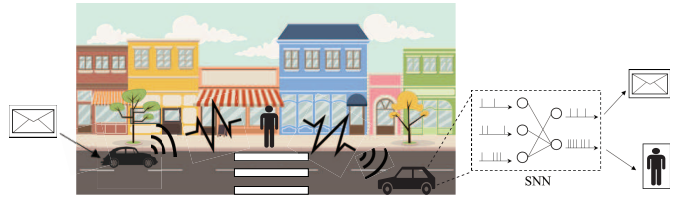}
        \vspace{1pt}
	\caption{This paper studies an ISAC system, in which the same IR (or UWB) signal is used for transmission and radar detection of the presence of a target. The key novel element is the use of neuromorphic computing at the ISAC receiver to simultaneously demodulate digital data and provide an online estimate of the presence or absence of the radar target. }
	\label{gmodel}
\end{figure}

As an example of an application that can benefit from ISAC, consider the inter-vehicle communication scenario in Fig.~\ref{gmodel}. In it, a car wishes to send a message to a second car, while also enabling the latter to detect the presence of a possible target, e.g., of a pedestrian. While conventional systems would use two separate radio resources for data transmission and radar detection, ISAC solutions reuse the same transmitted waveform for the dual role of carrier of digital information and radar signal \cite{jeong2014beamforming, liu2020joint}. A natural radio interface to serve this dual function is impulse radio (IR), also known as ultrawideband (UWB). In fact, IR encodes information in the timing of pulses, which can in turn be repurposed for radar detection \cite{nezirovic2009signal,zhang2009real}. 

This paper proposes to leverage the synergy between IR transmission and neuromorphic computing \cite{skatchkovsky2020end, chen2022neuromorphic} to realize efficient ISAC systems. Neuromorphic computing is an emerging computing technology that can efficiently process information encoded in binary signals known as \textit{spikes} \cite{davies2018loihi}. With the aim of reducing energy consumption and facilitating online and always-on operation on specialized hardware, as illustrated in Fig.~1, we introduce a neuromorphic ISAC (N-ISAC) receiver that leverages spiking neural network (SNN)-based processing to demodulate digital information and detect the radar signal.

\vspace{-6pt}
\subsection{Related Work}
\vspace{-3pt}
IR is widely used for low-power communications, e.g., in the IEEE 802.15.4z standard \cite{win1998impulse}, and is envisaged to be part of the ``gearbox'' of the physical layer of 6G systems \cite{fettweis20216g}. Among the key advantages of IR is the possibility to implement receivers that can operate using efficient analog or neuromorphic hardware \cite{cassidy2008impulse, shahshahani2015all, jouni20221}. The integration of IR and neuromorphic computing was investigated in \cite{skatchkovsky2020end, chen2022neuromorphic}, which proposed an end-to-end neuromorphic architecture for remote inference that replaces traditional digital blocks with SNNs as encoder and decoder. Hardware implementations includes \cite{cassidy2008impulse}, which introduced an IR-based communication protocol to convey digital packets between SNN chips (see also \cite{roth2022spike}); and reference \cite{shahshahani2015all}, which proposed an all-digital spike-based IR wireless transmission scheme for miniaturized biomedical applications.

One of the key research directions for ISAC systems is optimal waveform design \cite{9540344}. For instance, reference \cite{xiao2022integrated} studied ISAC based on delay alignment modulation techniques for Terahertz massive MIMO.
\vspace{-6pt}
\subsection{Main Contributions}
\vspace{-3pt}
To our best of knowledge, this is the first work to propose the implementation of ISAC via neuromorphic computing. 
The main contributions of this letter are summarized as follows. 
\begin{itemize}
\item We introduce the novel N-ISAC system illustrated in Fig.~\ref{model}, which exploits neuromorphic computing and IR to achieve simultaneous data transmission and target detection. The proposed SNN-based architecture demodulates digital data and detects the presence of a radar target in an online fashion;
\item We propose a supervised learning method for the design of the N-ISAC receiver in Fig.~\ref{model};
\item Numerical results are provided that demonstrate the advantages of the proposed N-ISAC system over conventional separate sensing and communication (SSAC) solutions. 
\end{itemize}

The remainder of the paper is organized as follows. Section \ref{system} presents the system model. The neuromorphic receiver processing is detailed in Section \ref{snn}, while Section \ref{exp} presents the experimental setting and results. Finally, Section \ref{con} concludes the paper.
\vspace{-10pt}

\section{System Model} \label{system}
\vspace{-3pt}
\begin{figure}[htp]
	\centering
	\includegraphics[width=3.5in]{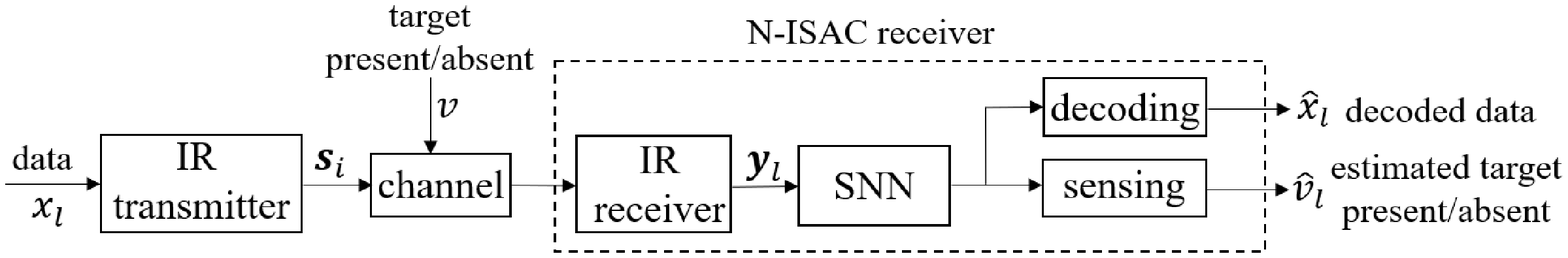}
        \vspace{1pt}
	\caption{N-ISAC: Digital data is transmitted by an IR transmitter via pulse-position modulation (PPM) as illustrated in Fig.~3; while the receiver simultaneously decodes digital data, and performs radar detection by means of an SNN, which can be efficiently implemented on neuromorphic hardware. }
	\label{model}
\end{figure} 
\vspace{-6pt}
As illustrated in Fig.~\ref{gmodel} and Fig.~\ref{model}, we consider an ISAC system in which digital communication and radar sensing leverage the same IR transmitted signal. In order to efficiently and simultaneously decode the digital data and detect the possible presence of a target at a known delay cell, the receiver processes the received signal via an SNN. Given the use of time-encoded information and neuromorphic computing, we refer to the proposed system as \emph{neuromorphic ISAC (N-ISAC)}.

\vspace{-10pt}
\subsection{IR Transmission}
\begin{figure}[htp]
	\centering
	\includegraphics[width=3.5in]{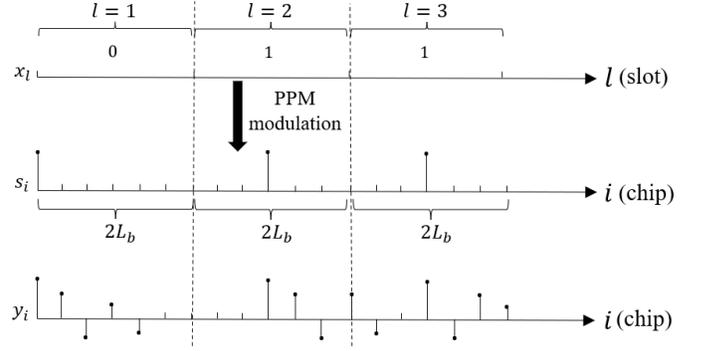}
        \vspace{1pt}
	\caption{PPM modulation: Each digital data $x_l\in\{0, 1\}$ at time slot $l$ is modulated via PPM, in which a spike at the first chip encodes $x_l=0$, while a spike at the $(L_b+1)$th chip encodes $x_l=1$. Due to the fading and multi-path effects of the channel, the PPM-modulated symbols $\{s_i\}$ are corrupted by the channel during transmission to produce the received symbols $\{y_i\}$. }
	\label{ppm}
\end{figure}

\vspace{-3pt}
The IR transmitter modulates and transmits digital data $x_l\in\{0,1\}$, for each discrete-time instant $l=1,\ldots,L$, using pulse position modulation (PPM) through a single antenna. Each bit $x_l$ is encoded within the $l$th time slot of duration $T$ seconds. Specifically, given an information bit sequence $\{x_1,\ldots, x_L\}$ of $L$ bits, the PPM-modulated signal is given as \cite{win2000ultra}
\begin{align}
	s(t)= \sum_{l=1}^{L} \phi\big(t-(l-1)T-x_lL_bT_c\big),
	\label{modualation}
\end{align}
where $\phi(t)$ is the pulse waveform of bandwidth $1/T_c$; and $T_c$ is the chip time. The number of chips within each slot is $T/T_c=2L_b$, where integer $L_b \geq 1$ is referred to as the \emph{bandwidth expansion factor}. Accordingly, as illustrated in Fig.~\ref{ppm}, in each slot $l$, bit $x_l$ is encoded by a pulse on the first chip when $x_l=0$, and by a pulse in the middle of the slot, i.e., at the $L_b+1$th chip, when $x_l = 1$. The energy of the pulse waveform $\phi(t)$ is constrained to be smaller than a given value $E_b$, i.e., $\int_{-\infty}^{+\infty}\phi^2(t) \leq E_b$.

\vspace{-7pt}

\subsection{Multipath Channel and Radar Target}
\vspace{-3pt}
\label{multipath}
The modulated signal $s(t)$ is transmitted over a multi-path fading channel to the receiver. The multi-path channel depends on the possible presence of a radar target, as well as on scatterers located between the transmitter and receiver that act as \emph{clutter} for radar processing. We use the binary variable $v$ to indicate the presence or absence of the radar target,  identified respectively by setting $v=1$ and $v=0$. Specifically, following a standard radar model (see, e.g., \cite{shnidman1999generalized}), we assume that a target may or may not be present at a radar cell corresponding to a known propagation delay $\tau_0$. Accordingly, the continuous-time channel response can be expressed as
\begin{align}
	h(t)= v \beta_{0} g(t-\tau_0)+ \sum_{c=1}^{N_c}\beta_{c} g(t-\tau_{c}),
	\label{h}
\end{align}
where $g(t)=\phi(-t)$ is the receiver filter response, which is matched to the transmitted waveform $\phi(t)$; the amplitude $\beta_0$ of the target follows a complex Gaussian distribution with power $\sigma_0^2$, i.e., $\beta_0 \sim \mathcal{CN}(0,\sigma_0^2)$; and the amplitudes $\{\beta_c\}_{c=1}^{N_c}$ of the $N_c$ clutter components are independent with uniform phases and Weibull absolute values having shape parameter $\kappa\in[0.25,2]$ and scale parameter $\lambda\in(0,\infty)$ \cite{shnidman1999generalized}. With $\kappa=2$, the amplitudes $\{\beta_c\}_{c=1}^{N_C}$ are complex Gaussian random variables \cite{shnidman1999generalized}. The amplitudes $\beta_0$ and $\{\beta_c\}_{c=1}^{N_c}$ and delays $\{\tau_c\}_{c=1}^{N_c}$ are assumed to be unknown.

We note that the model \eqref{h} could be generalized to allow for a time-varying presence/absence pattern for the target, in the sense that the variable $v$ could change over the time slot index $l$. We will not consider this situation in this paper, although the extension is straightforward.

\vspace{-6pt}
\subsection{Receiver Processing}
\vspace{-3pt}

The signal obtained by the single-antenna receiver is given by 
\begin{align}
	y(t)= s(t)*h(t)+z(t),
	\label{re}
\end{align}
where the channel response $h(t)$ is given in \eqref{h}; $z(t)$ is additive white Gaussian noise with power spectral density $N_0$; and  ``$*$'' denotes the convolution operation. The receiver samples the received signal $y(t)$ at the chip rate $1/T_c$, yielding the discrete-time signal $y_{i}=y(iT_c)$, for $i=1,2,\ldots,2L_bL$. For each slot time $l$, we collect the $2L_b$ samples $\mv y_l=\{y_i\}_{i\in\mathcal{I}_l}$, with $i\in\mathcal{I}_l=\{2(l-1)L_b+1,\ldots, 2lL_b\}$. Assuming a maximum delay spread $T_h$, upon sampling, the effective discrete-time channel has $L_h=T_h/T_c$ taps. Accordingly, the channel between the transmitter and the receiver is described by the $L_h\times1$ vector $\mv h=[h(0), h(T_c),\ldots,h((L_h-1)T_c)]^T$. Note that, if $L_h<L_b$, there is no interference between pulses transmitted in successive slots; while, otherwise, transmission of the pulse encoding the $l$th bit $x_l$ may interfere with the pulse encoding the following bits $\{x_{l^{\prime}}\}_{l^{\prime}>l}$.

Using \eqref{re}, the received discrete-time signal $y_{i}$ for the $i$th chip can be expressed as
\begin{align}
	 y_{i}= \mv h^T \mv s_{i} + z_{i},
	\label{y_l}
\end{align}
where we have defined the $L_h \times 1$ vector $\mv s_{i}=[s_{i}, s_{i-1}, \ldots, \\ s_{i-L_h+1}]^T$, with $s_{i}=s(iT_c)$ and $s_{i}=0$ if $i\leq0$; and $z_{i}=z_n(iT_c)\sim \mathcal{CN}(0,N_0B)$ are the noise samples, where $B=1/T_c$ is the pulse bandwidth. By the assumed PPM scheme, for each slot $l$, we have all zero samples $\{s_i\}_{i\in\mathcal{I}_l}$ except for one sample, namely $s_{2(l-1)L_b+1}=1$ if $x_l=0$ and $s_{2(l-1/2)L_b+1}=1$ if $x_l=1$. 

The receiver feeds the received samples $\{y_i\}_{i\in\mathcal{I}}$ to the SNN successively for each slot $l$. To this end, we collect the received samples $\{y_{i}\}_{i\in\mathcal{I}}$ corresponding to the time slot $l$ into a $2L_b\times 1$ vector $\mv y_{l}$. Since $\mv y_l$ is generally complex, we define the $4L_b\times1$ real-value vector 
\begin{align}
\bar{\mv y}_{l}=[\mathfrak{R}(\mv y_{l})^T, ~\mathfrak{S}(\mv y_{l})^T]^T, \label{spikingy}
\end{align}
with $\mathfrak{R}(\cdot)$ and $\mathfrak{S}(\cdot)$ being the element-wise real and imaginary parts of the input vector, respectively. The received signal $\bar{\mv y}_{l}$ is input to the SNN at each time step $l$. 

As described in the next section, an SNN is a recurrent discrete-time model that operates along the slot index $l$. It takes as input the $2L_b$ real samples $\bar{\mv y}_l$, with one output neuron producing the decoding output  $\hat{x}_l \in \{0,1\}$, which provides an estimate of the data symbol $x_l$, and the other output neuron producing the sensing output $\hat{v}_l$, which estimates the binary variable $v_l\in \{0,1\}$ representing the presence or absence of the target. Note that the detection decision $\hat{v}_l$ about the target varies over time as more information is acquired by the receiver.

\vspace{-5pt}
\section{Neuromorphic Receiver Processing} 
\vspace{-3pt}
\label{snn}
In this section, we first describe the SNN model used at the receiver, and then we detail the proposed data-aided optimization of the model parameters of the SNN.
\vspace{-6pt}


\subsection{Spiking Neural Network Model}
\vspace{-3pt}
An SNN is specified as a network connecting a set of spiking neurons via an arbitrary directed graph, in which a directed edge represents a synapse. To model the spiking neurons, we adopt the standard discrete-time spike response model (SRM), in which each spiking neuron outputs a binary signal $b_{l}\in\{0,1\}$, with ``1'' representing the emission of a spike and ``0'' an idle neuron at time step $l=1,\ldots,L$. Each neuron $k$ maintains an internal analog state variable $o_{k,l}$, known as the \emph{membrane potential}, over time step $l$.

The membrane potential $o_{k,l}$ evolves over time $l=1,2,\ldots$ as a function of the spikes that neuron $k$ receives from the neurons that have synapses ending at neuron $k$. The neuron spikes when the membrane potential crosses a threshold, after which the membrane potential is reduced below the threshold. Mathematically, the membrane potential $o_{k,l}$ is defined by the sum of filtered contributions from incoming spikes and from the neuron $k$'s own past outputs. Accordingly, the evolution of the membrane potential is modelled as 
\begin{align}
	o_{k,l}=\sum_{j\in\mathcal{P}_{k}}w_{k,j}\cdot(\alpha_l * b_{k,l})+\beta_l * b_{k,l},
	\label{potential}
\end{align}
where $\mathcal{P}_k$ is the set of neurons with synapses ending at neuron $k$; $w_{k,j}$ is the synaptic weight between neuron $j\in\mathcal{P}_{k}$ and neuron $k$; $\alpha_l$ represents the synaptic response to a spike from a neuron $j\in\mathcal{P}_{k}$; $\beta_l$ describes the synaptic response to the spike emitted by neuron $k$ itself; and ``$*$'' is the convolution operator. Typical choices for synaptic spike responses include the first-order feedback filter $\beta_l=\exp(-l/\tau_{\rm ref})$, and the second-order synaptic filer $\alpha_l=\exp(-l/{\tau_{\rm mem}})-\exp(-l/{\tau_{\rm syn}})$, for $l=1,2,\ldots$, with finite positive constants $\tau_{\rm ref}$, $\tau_{\rm mem}$ and $\tau_{\rm syn}$. We refer to \cite{skatchkovsky2021spiking, jang2019introduction} for further details.

Neuron $k$ outputs a spike at time step $l$ if its membrane potential $o_{k,l}$ passes some fixed threshold $\vartheta$. Accordingly, the output of neuron $k$ can be expressed as
 \begin{align}
	b_{k,l}=\Theta(o_{k,l}-\vartheta),
	\label{spike}
\end{align}
where $\Theta(\cdot)$ is the Heaviside step function ($\Theta(x)=1$ if $x>0$ and $\Theta(x)=0$ otherwise). We denote as $\mv \theta$ the vector of all parameters $\{w_{k,j}\}$ for the SNN of the receiver.

The SNN at the receiver has two read-out neurons producing decisions $\hat{x}_l$ and $\hat{v}_l$. Denoting as $k^c$ the index of the communication readout neuron, and as $k^s$ the index of the radar readout neuron, by \eqref{spike} we have the estimates $\hat{x}_l=b_{k^c,l}$ and $\hat{v}_l=b_{k^s,l}$. 

\subsection{Optimization}
To optimize the model parameters of the SNN, we adopt a supervised learning approach, and we assume the availability of a dataset $\mathcal{D}$ consisting of examples of the form $(\bar{\mv y}, \mv x, v)$. Each example $(\bar{\mv y}, \mv x, v)$ contains the sequence of $L$ transmitted bits $\mv x= \{x_l\}_{l=1}^L$ with $x_l\in\{0,1\}$; the corresponding $L$ received samples $\bar{\mv y}=\{\bar{\mv y}_l\}_{l=1}^L$ obtained via (5); and the binary value $v$ indicating the presence or absence of the target, which is assumed to be the same across $L$ samples. For each example $(\bar{\mv y}, \mv x, v)$, at each $l$th step, the received signal $\bar{\mv y}_l$ serves as input to the SNN, while $x_l$ and $v$ are ground-truth labels for data decoding and target detection, respectively.

To define the training criterion, we evaluate the performance of communication and sensing via two separate cross-entropy losses. To this end, defining the sigmoid function $\sigma(x)=(1+e^{-x})^{-1}$, we write the probability $p^c_l(\mv \theta)=\sigma(o_{k^c,l})$ of the communication readout neuron to produce the estimate $\hat{x}_l=1$; and the probability $p^s_l(\mv \theta)=\sigma(o_{k^s,l})$ of the radar readout neuron to produce the estimate $\hat{v}_l=1$. They are both functions of the model parameters $\mv \theta$. The training loss related to data decoding is then measured by the cross-entropy loss
\begin{align}
    \ell^c_l(\mv \theta) =-x_l \log(p_l^c(\mv \theta))-(1-x_l)\log(1-p_l^c(\mv \theta)),
	\label{dloss}
\end{align}
while the training loss for sensing is similarly defined as 
\begin{align}
    \ell^s_l(\mv \theta) =-v \log(p_l^s(\mv \theta))-(1-v)\log(1-p_l^s(\mv \theta)).
	\label{sloss}
\end{align}
The overall training losses for each data point are defined as the sums over time $L^c(\mv \theta)=\sum_{l=1}^L\ell_l^c(\mv \theta)$ and $L^s(\mv \theta)=\sum_{l=1}^L\ell_l^s(\mv \theta)$. For the considered SNN architecture, we adopt the weighted sum
\begin{align}
    L^{ISAC}(\mv \theta) =  \beta L^c(\mv \theta) + (1-\beta) L^s(\mv \theta)
	\label{loss}
\end{align}
of the communication and sensing losses, where $\beta\in[0,1]$ is a weight factor determining the relative priority between the two losses. The model parameters $\mv \theta$ are updated based on stochastic gradient descent (SGD). To this end, we address the nondifferentiable threshold activation \eqref{spike} via surrogate gradient \cite{neftci2019surrogate} by replacing the Heaviside step function in \eqref{spike} with a differentiable surrogate function, namely the sigmoid $\sigma(x)$, during training.
\vspace{-5pt}

\section{Experiments} 
\vspace{-3pt}
\label{exp}
In this section, we provide experimental results on the proposed N-ISAC system. We start by describing setting and benchmarks.
\vspace{-6pt}
\subsection{Setting}
\vspace{-3pt}
The system operates over $L=80$ slots. The information bits $\{x_l\}_{l=1}^L$ and target $v$ are i.i.d. generated as $1/2$-Bernoulli variables, i.e., $x_l \sim \text{Bern}(0.5)$, and $v\sim \text{Bern}(0.5)$. The target delay is set to $\tau_0=0$, and the amplitudes $\{\beta_c\}_{c=0}^{N_c}$ are all i.i.d. $\beta_c\sim \mathcal{CN}(0,1)$ variables, while the delays $\{\tau_c\}_{c=1}^{N_c}$ are uniformly distributed between 0 and $4T_c$. We set the number of clutter paths in the channel \eqref{h} as $N_c=5$. We generate $60,000$ training examples and $10,000$ test examples as described in Section \ref{multipath}. The SNN has a fully connected architecture with a single hidden layer, with $4L_b$ input neurons and $6$ or $10$ hidden neurons. The average signal-to-noise ratio (SNR) is defined as $\mathrm{SNR}=\mathbb{E}[\|\mv h\|^2] E_b/(N_0B)$, where $\mathbb{E}[\|\mv h\|^2]$ is the average squared norm of the channel. We set the SNR to $10$dB.

\vspace{-8pt}
\subsection{Benchmark}
\vspace{-2pt}
For comparison, we consider a conventional SSAC scheme. SSAC divides the $L$ slots into $\lceil \alpha L \rceil$ slots used to transmit information, and $L-\lceil \alpha L \rceil$ slots used for radar sensing, where $\alpha\in [0,1]$. The transmitted signal for radar sensing is given by the PPM waveform \eqref{modualation} with bits fixed to $\{x_l=1\}_{l=\lceil \alpha L \rceil+1}^L$. For SSAC, two SNNs are implemented at the receiver, one performing data decoding for the first $\lceil \alpha L \rceil$ slots, and the other responsible for radar sensing in the rest of the time slots. The architecture of both SNNs is as described above, with each SNN having $6$ hidden neurons.
\vspace{-6pt}
\subsection{Results}
\vspace{-3pt}
We adopt the following performance metrics for data transmission and radar sensing:
\begin{itemize}
    \item \emph{Normalized test throughput}, i.e., the ratio $\mathbb{E}[L_{\rm succ}]/L$ of the average number $L_{\rm succ}<L$ of correctly decoded bits over the total number of time slots $L$;
    \item \emph{Radar test detection error}, i.e., the probability that the sensing decision $\hat{v}$ is not correctly taken upon processing all time slots. The final decision $\hat{v}$ for each example is made based on the majority rule, predicting the presence of a target $(\hat{v}=1)$ if the SNN produces the decision $\hat{v}_l=1$ for a majority of the time slots $l=1,\ldots,L$.
\end{itemize}

\begin{figure}[t!]
	\centering
	\includegraphics[width=3.2in]{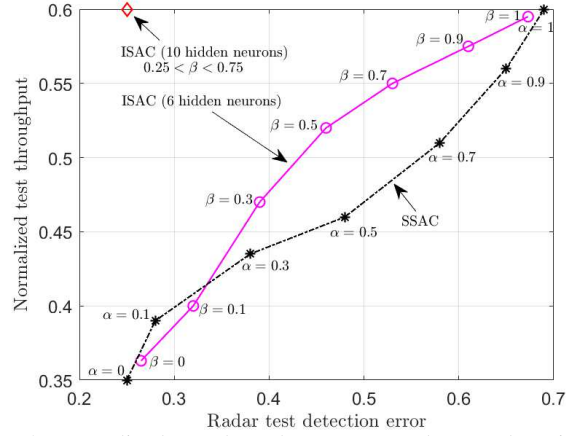}
        \vspace{5pt}
	\caption{Normalized test throughput versus radar test detection error for ISAC and SSAC $(L_b=1$$)$.}
	\label{c1}
\end{figure} 

\begin{figure}[t!]
	\begin{center}
		\subfigure{\includegraphics[width=3in]{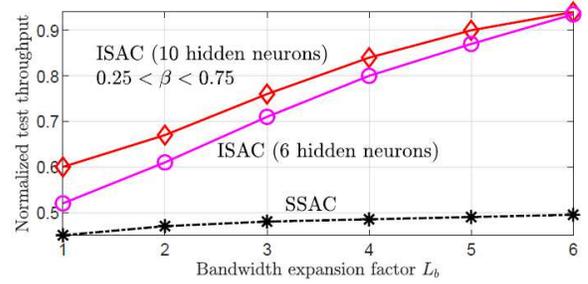}}\hfill
		\subfigure{\includegraphics[width=3in]{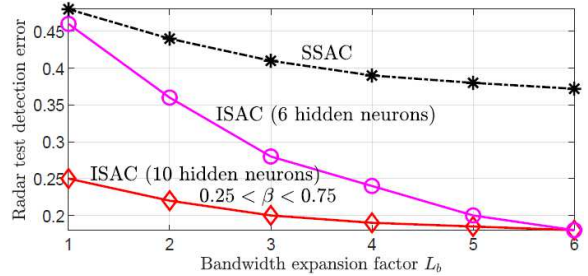}}
	\end{center}
        \vspace{-2pt}
	\caption{(Top) Normalized test throughput and (Bottom) radar test detection error versus bandwidth expansion factor $L_b$ for ISAC and SSAC $(\alpha=\beta=0.5$$)$. \label{c2}}
\end{figure}

We demonstrate the normalized test throughput versus the radar test detection error for ISAC and SSAC when there is no bandwidth expansion, e.g., for $L_b=1$, in Fig.~\ref{c1}. For the ISAC scheme, we vary $\beta$ in the loss \eqref{loss}, while we change the fraction $\alpha$ for SSAC. As $\beta$ increases, more priority is given by ISAC to communication over radar detection; and, similarly, as $\alpha$ increases, SSAC assigns more slots to communications. The performance of ISAC with an SNN having $10$ hidden neurons is essentially independent of $\beta$ for any $0.25<\beta<0.75$. A first observation is that, for SSAC, there is a trade-off between communication and sensing performance levels caused by the slot allocation. A similar trade-off is also observed for ISAC when using an SNN with $6$ hidden neurons. This is due to the limited capacity of the shared common hidden layer of the SNN. In contrast, when $10$ hidden neurons are available at the SNN, ISAC is seen to optimize both data decoding and target sensing performance, obtaining significant gains over SSAC.

In Fig.~\ref{c2}, we demonstrate the normalized test throughput and radar test detection error versus the bandwidth expansion factor $L_b$ for ISAC and SSAC by setting $\alpha=\beta=0.5$. Both ISAC and SSAC are observed to benefit from the bandwidth expansion due to the reduced inter-slot interference. However, the throughput achievable by SSAC is bounded by $\alpha=0.5$, while ISAC can obtain larger throughputs; and a similar behavior is observed also in terms of the probability of detection error. Furthermore, when the bandwidth expansion $L_b$ is sufficiently large, e.g., when $L_b=6$, ISAC with $6$ hidden neurons obtains a similar performance as that with $10$ neurons, which indicates a larger bandwidth simplifies receiver design in terms of SNN's capacity. 

\begin{figure}[t!]
	\centering
	\includegraphics[width=3.5in]{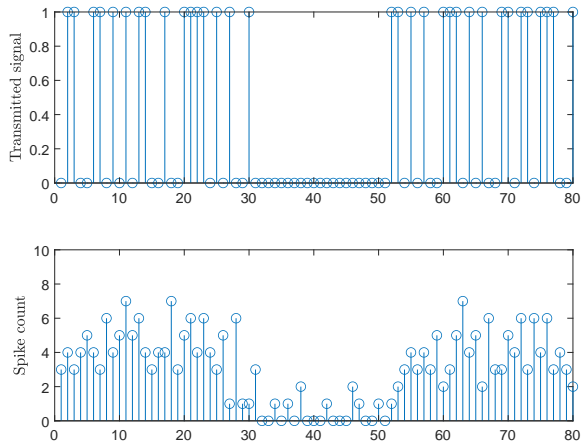}
        \vspace{0pt}
	\caption{Top: Transmitted signal consisting of two frames in which the transmitter is active separated by an idle frame. Bottom: Corresponding spike count for the SNN ($\beta=0.5$, SNR=10dB).}
	\label{time}
\vspace{-2pt}
\end{figure} 
Finally, Fig.~\ref{time} illustrates how the SNN receiver can leverage the temporal sparsity of the IR signals to enhance energy efficiency. In this regard, we recall that energy consumption in an SNN is essentially proportional to the number of spikes produced by the SNN, given extremely low idle energy of neuromorphic chips \cite{davies2018loihi}.  The top panel shows the transmitted IR signal consisting of two frames of transmitted signals with $L_b=1$, separated by an idle frame of duration of $20$ slots. We observe that in the idle frame, the spike count is significantly reduced, showing that the neuromorphic receiver can adjust its energy consumption to the activity level of the transmitter.

\vspace{-5pt}
\section{Conclusions}
\vspace{-3pt}
\label{con}
In this paper, we have proposed a novel ISAC solution that leverages the synergy of neuromorphic computing and IR transmission for both data transmission and radar detection. Data are encoded on an IR waveform that is used for the dual purpose of data decoding and target sensing. A SNN-based receiver architecture is considered to estimate data and the presence/absence of a target. Experiments have demonstrated the advantage of the proposed neuromorphic ISAC over conventional separate sensing and communications in terms of normalized test throughput and radar test detection error. We have also highlighted the capacity of the neuromorphic receiver to adapt its computing energy consumption to the activity level of the transmitter.
\vspace{-5pt}

\small{\bibliographystyle{IEEEtran}
\bibliography{references}}

\end{document}